\documentclass[sigconf]{acmart}
\usepackage{enumitem}
\usepackage{multirow}
\AtBeginDocument{%
  }

\setcopyright{acmlicensed}
\copyrightyear{2018}
\acmYear{2018}
\acmDOI{XXXXXXX.XXXXXXX}
\acmConference[Conference acronym 'XX]{Make sure to enter the correct
  conference title from your rights confirmation email}{June 03--05,
  2018}{Woodstock, NY}
\acmISBN{978-1-4503-XXXX-X/2018/06}

\renewcommand\footnotetextcopyrightpermission[1]{}

\setcopyright{none}

\begin{document}
\title{AmbiGraph-Eval: Can LLMs Effectively Handle Ambiguous Graph Queries?}

\author{Yuchen Tian}
\affiliation{
  \institution{Hong Kong Baptist University}
  \country{Hong Kong, China}
}
\email{yctian@comp.hkbu.edu.hk}

\author{Kaixin Li}
\affiliation{
  \institution{National University of Singapore}
  \country{Singapore}
}
\email{likaixin@u.nus.edu}

\author{Hao Chen}
\affiliation{
  \institution{BIFOLD \& TU Berlin}
  \country{Berlin, Germany}
}
\email{hao.chen@tu-berlin.de}

\author{Ziyang Luo}
\affiliation{
  \institution{Hong Kong Baptist University}
  \country{Hong Kong, China}
}
\email{cszyluo@comp.hkbu.edu.hk}

\author{Hongzhan Lin}
\affiliation{
  \institution{Hong Kong Baptist University}
  \country{Hong Kong, China}
}
\email{cshzlin@comp.hkbu.edu.hk}
\author{Sebastian Schelter}
\affiliation{
  \institution{BIFOLD \& TU Berlin}
  \country{Berlin, Germany}
}
\email{schelter@tu-berlin.de}

\author{Lun Du}
\affiliation{
  \institution{Ant Group}
  \country{Beijing, China}
}
\email{dulun.dl@antgroup.com}

\author{Jing Ma}
\affiliation{
  \institution{Hong Kong Baptist University}
  \country{Hong Kong, China}
}
\email{majing@comp.hkbu.edu.hk}

\renewcommand{\shortauthors}{Trovato et al.}

\begin{abstract}
Large Language Models (LLMs) have recently demonstrated strong capabilities in translating natural language into database queries, especially when dealing with complex graph-structured data. However, real-world queries often contain inherent ambiguities, and the interconnected nature of graph structures can amplify these challenges, leading to unintended or incorrect query results. 
To systematically evaluate LLMs on this front, we propose a taxonomy of graph-query ambiguities, comprising three primary types: \textit{Attribute Ambiguity}, \textit{Relationship Ambiguity}, and \textit{Attribute-Relationship Ambiguity}, each subdivided into \textit{Same-Entity} and \textit{Cross-Entity} scenarios.
 We introduce \textbf{AmbiGraph-Eval}, a novel benchmark of real-world ambiguous queries paired with expert-verified graph query answers. Evaluating 9 representative LLMs shows that even top models struggle with ambiguous graph queries. Our findings reveal a critical gap in ambiguity handling and motivate future work on specialized resolution techniques.
\end{abstract}

\maketitle

\section{Introduction}
\label{sec:intro}
Semantic parsing translates natural language into formal query languages (e.g., SQL or Cypher), enabling intuitive interaction between users and databases~\citep{deng2022recent}. However, natural language is inherently ambiguous and often admits multiple interpretations. In contrast, query languages require precise and unambiguous expressions to operate correctly. While recent studies have addressed ambiguity in SQL queries over tabular data~\citep{he2024text2analysis, bhaskar2023benchmarking}, graph databases—commonly used in domains such as social media and finance—introduce additional complexity due to their rich, interconnected schema. These structural intricacies further amplify the challenge of mapping user intent to accurate graph queries, often leading to unintended results or inefficient data retrieval~\cite{10.1145/3701716.3715450}.

\begin{figure}[t]
\centering
\includegraphics[width=0.47\textwidth]{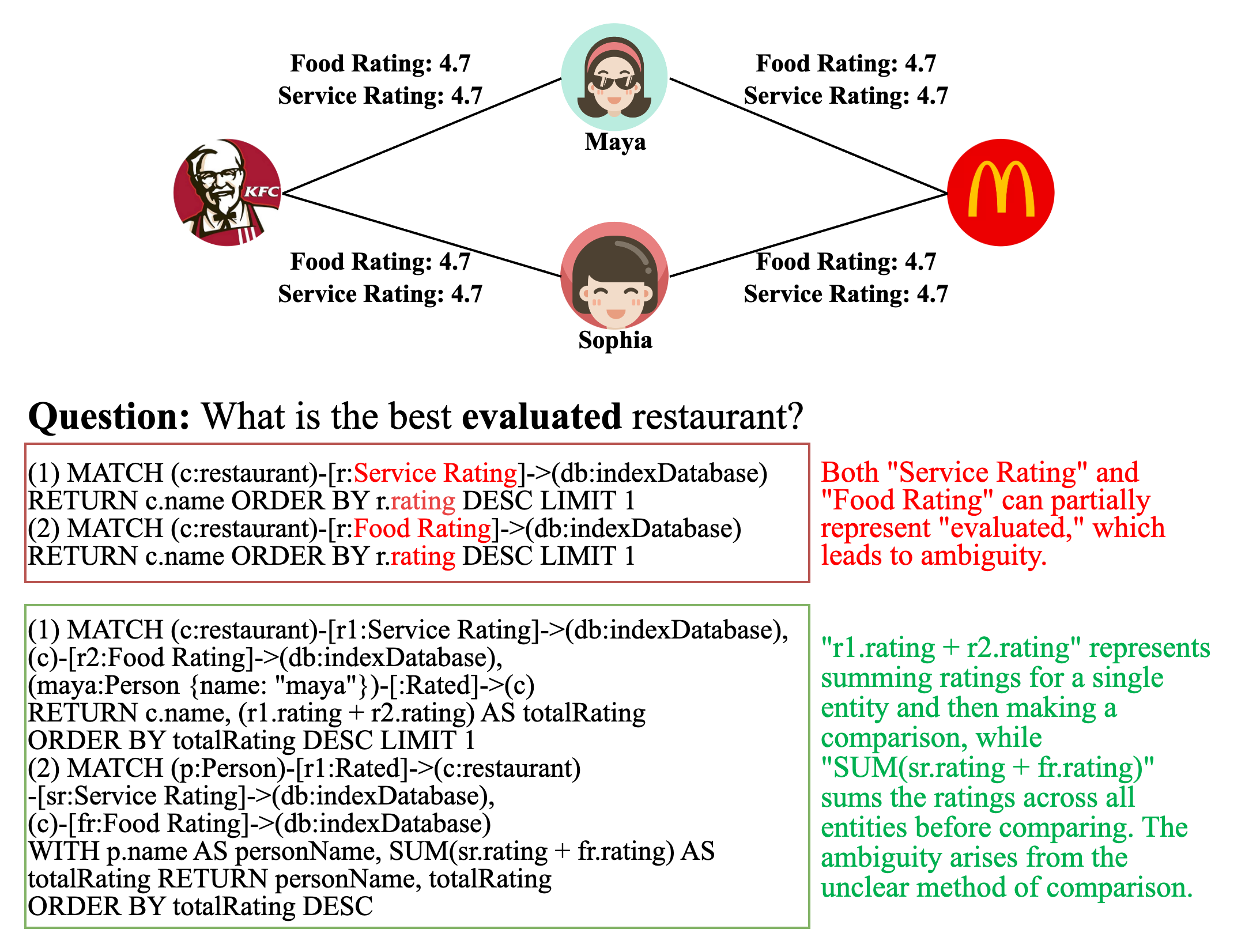}
\caption{
Ambiguous example in text-to-CQL task. Blue font denotes the problematic question span and red font means the “plausible” relationship. 
}
\label{fig:example_ambiguity}
\end{figure}

In particular, natural language queries involving nodes and relationships in graphs often give rise to multiple interpretations, owing to the structural richness and diversity of graph data. For example, as illustrated in Figure~\ref{fig:example_ambiguity}, the query ``best evaluated restaurant'' can yield different results depending on how the evaluation criteria are interpreted. Because the graph does not contain a relationship or attribute explicitly labeled as ``evaluated'', the system resorts to fuzzy matching with semantically related properties, such as ``Service Rating'' and ``Food Rating'', resulting in the candidate answers highlighted in the red box.  Furthermore, the comparison strategy itself is ambiguous: the system might compare individual user ratings (e.g., Maya’s scores) or aggregate ratings across all users, leading to alternative results shown in the green box. If such ambiguities are not properly resolved, the generated query may deviate from the user’s intent, resulting in inefficient or irrelevant data retrieval and computation.

The presence of such ambiguities in interactive systems introduces significant risks. If semantic parsing techniques fail to detect and manage these ambiguities, the system may generate queries that deviate entirely from the user's intended request. This misalignment can lead to unnecessary data retrieval and computation, wasting valuable resources and time. In scenarios where real-time decision-making or resource efficiency is critical, these missteps can severely impact system performance, increasing operational costs and reducing overall effectiveness.

With the rapid advancement of large language models (LLMs), LLM-based semantic parsing~\citep{drozdov2022compositional} has shown strong potential in handling complex and ambiguous user queries. These models leverage their linguistic knowledge to interpret nuanced inputs and confirm the intended query through interactive clarification with the user. However, despite these advances, current LLMs exhibit a notable limitation: \textit{self-preference bias} \cite{ye2024justice}. Because LLMs are trained using human feedback, they may internalize the preferences of annotators, causing them to favor certain interpretations over others. This bias can lead to systematic misalignment between the model’s output and the user’s actual intent, especially in cases where multiple interpretations are equally valid.

In this work, we evaluate the ability of current LLMs and related techniques to handle ambiguity in graph query generation. Specifically, we: (1) systematically categorize the types of ambiguity that arise in graph queries, (2) introduce a curated dataset comprising ambiguous user queries and corresponding graph data with human-labeled graph queries (Cypher), and (3) conduct a comprehensive evaluation of existing LLMs and relevant techniques to assess their ability to resolve these ambiguities effectively.

We introduce the concept of ambiguity in graph database queries, categorizing it into three primary types based on real-world data: \textbf{Attribute ambiguity}, arising from unclear attribute definitions; \textbf{Relationship ambiguity}, stemming from uncertainty in relationships or their attributes; and \textbf{Attribute-Relationship ambiguity}, where both attributes and relationships contribute to the uncertainty. Each type can be further divided into \textbf{Same Entity ambiguity}, where ambiguity pertains to a single entity, and \textbf{Cross Entity ambiguity}, where uncertainty arises from interactions across multiple entities.

To evaluate model performance, we propose \textit{AmbiGraph-Eval}, a benchmark specifically designed for assessing the ability of LLMs to handle ambiguity in graph queries. \textit{AmbiGraph-Eval} comprises 560 carefully curated ambiguous queries and corresponding graph database samples. Using this benchmark, we test 9 LLM models, analyzing their ability to resolve graph ambiguities, identifying model errors, and exploring strategies for improvement. Our findings indicate that reasoning capabilities offer limited benefits in this task. Instead, the models' understanding of graph ambiguity and their mastery of graph database query syntax play a more critical role. Through this study, we aim to advance the development of complex query generation and ambiguity resolution in graph query semantic parsing tasks.

Our contributions can be summarized as follows:
\begin{itemize}
\item Based on real-world data observations, we introduced the concept of ambiguity in graph queries, categorizing it into three main types. This addresses a significant gap in the research on ambiguity in Text-to-Graph query tasks in Section  \ref{sec:definatin_classfication}.

\item We released \textit{AmbiGraph-Eval}, a carefully curated dataset collected through a two-stage review process, designed to evaluate graph query ambiguity. Alongside this, we introduced a novel metric, i.e., \textit{Ambiguity-Resolved Execution Accuracy (AREA)} metric, for measuring model performance in resolving graph query ambiguity in Section \ref{sec:bench}.

\item We evaluated 9 LLMs using zero-shot strategy, providing a thorough analysis of the performance in Section \ref{sec:experiments}. 
\end{itemize}

\section{Ambiguity Categorization}
\label{sec:definatin_classfication}
In this section, we introduce our systematic categorization of ambiguity in graph query parsing, based on analysis of 22 graph databases. By examining node and edge attributes, along with data heterogeneity across sources, we identified three primary types of ambiguity: \textbf{Attribute ambiguity}, \textbf{Relationship ambiguity}, and \textbf{Attribute-Relationship ambiguity}. Each category is further divided into \textbf{Same Entity ambiguity} and \textbf{Cross Entity ambiguity}, ensuring comprehensive coverage of potential ambiguities in graph databases.

\textbf{Attribute ambiguity} occurs when a user queries an entity’s attribute, but multiple plausible attributes in the graph could match the query. This often arises in graphs with different data sources or representations of similar information. For example, querying the citation count of a paper titled ``CodeHalu'' may yield multiple counts from sources like Google Scholar or Semantic Scholar, creating uncertainty about which count the user intends. Ambiguity can also occur in comparisons, such as "Which paper has the highest citation count?" if the citation source is not specified. We divide this ambiguity into two sub-types:  \textbf{Same Entity ambiguity}: When a single entity has multiple attributes that could match the user’s intent, creating uncertainty about the correct attribute; \textbf{Cross Entity ambiguity}: When comparing multiple entities without clear specification of which attribute should be used, making the comparison unclear.

\textbf{Relationship ambiguity} arises when a query involves an entity’s relationships, but the specific relationship required to resolve the query is unclear. This may happen when multiple relationships could potentially satisfy the query, or when the relationships themselves are complex. For instance, asking for an architect's ``contributions'' could refer to `design contributions' or `project contributions', creating ambiguity in how the query should be answered. Similar uncertainty can arise in comparative queries, such as ``Who is the best doctor according to patient reviews?'' where the review criteria (e.g., score vs. rating) are unclear. Similarly, this ambiguity can be divided into two sub-types: \textbf{Same Entity ambiguity}: When a single entity has multiple relationships, leading to uncertainty about which relationship is relevant to the query; \textbf{Cross Entity ambiguity}:  When comparing entities based on relationships, but the criteria or standards for comparison are unclear.

\textbf{Attribute-Relationship ambiguity} occurs when both an entity’s attributes and relationships are involved, making it unclear whether the information should be derived from the entity’s attributes or its relationships. For example, the query ``What is the rating for the movie Inception?'' could refer to an internal attribute (e.g., `IMDb rating') or an external relationship (e.g., reviews from external sources). Similarly, in the query ``Which city hall has the highest evaluation?'', the ambiguity arises from whether the focus is on the internal rating or external reviews.  \textbf{Same Entity ambiguity}: When the boundary between an entity’s attributes and its relationships is unclear, it causes uncertainty in how to answer the query. \textbf{Cross Entity ambiguity}: When comparing multiple entities where both attributes and relationships could influence the answer, it is unclear which to prioritize.

\begin{figure*}[t]
    \centering
    \includegraphics[width=1\textwidth]{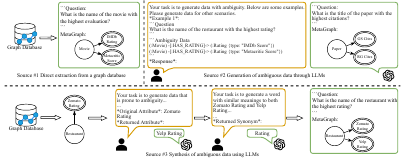}
    \caption{The construction process of \textit{AmbiGraph-Eval}. It is divided into two stages: data collection and human review.}
    \label{fig:ambigraph_generation_process}
\end{figure*}

\begin{table}
\caption{Detailed statistics of categories, subcategories, and sample quantities in \textit{AmbiGraph-Eval}.}
  \centering
  \resizebox{0.5\textwidth}{!}{ 
  \setlength{\tabcolsep}{10pt} 
  \begin{tabular}{cccc}
    \toprule
    \textbf{Category} & \textbf{Sub-Category} & \textbf{Samples} & \textbf{Ambig/Unambig} \\
    \midrule
    \multirow{2}{*}{Attribute} 
     & same-entity & 86 & 50/36 \\
     & cross-entity & 101 & 65/36 \\
    \midrule
    \multirow{2}{*}{Relationship} 
     & same-entity & 86 & 49/37 \\
     & cross-entity & 101 & 56/45 \\
    \midrule
    \multirow{2}{*}{Attribute-Relationship} 
     & same-entity & 94 & 57/37 \\
     & cross-entity & 92 & 59/33 \\
    \bottomrule
  \end{tabular}
  } \label{table:Detailed_statistics}
\end{table}

\section{AMBIGRAPH-EVAL BENCHMARK}
\label{sec:bench}
\textit{AmbiGraph-Eval} is a benchmark designed to evaluate large language models (LLMs) on their ability to generate syntactically correct and semantically appropriate logical queries (e.g., Cypher) in response to ambiguous natural language inputs grounded in graph data. Each query may correspond to multiple valid interpretations, challenging models to resolve ambiguity effectively.

The dataset is constructed in two stages: \textit{data collection} and \textit{human review}, as shown in Figure \ref{fig:ambigraph_generation_process}. During data collection, ambiguous examples are sourced through three complementary approaches:
\begin{itemize}[leftmargin=*, itemsep=0pt, topsep=0pt]
    \item Direct extraction from graph databases: identifying inherently ambiguous structures or incomplete information and manually creating ambiguous questions;
    \item Synthesis from unambiguous data using LLMs:  transforming clear graph data into ambiguous forms by introducing semantic overlaps in attributes and relationships;
    \item Full generation using LLMs: prompting models to create new ambiguous cases by generalizing from validated seed examples.
\end{itemize}
These methods cover a broad range of real and synthetic ambiguity types, including attribute ambiguity, relationship ambiguity, and their combination.

In the human review stage, experts rigorously examine the questions, graph structures, and reference answers to ensure ambiguity is genuine and answers are valid. The review involves multiple rounds of validation and voting, where any disputed item is removed. This process results in 560 high-quality ambiguous examples, with statistics reported in Table~\ref{table:Detailed_statistics}.

\begin{table*}[t]
\caption{Evaluation results of 9 models on \textit{AmbiGraph-Eval}. SE denotes same-entity. CE denotes cross-entity. Results are measured by $P_{\text{AREA}}$ (\%)}
\resizebox{\textwidth}{!}{%
  \begin{tabular}{lccccccccccccccccc}
    \toprule
    \multirow{2.3}{*}{\textbf{Model}} & \multicolumn{3}{c}{\textbf{Attribute }} & & \multicolumn{3}{c}{\textbf{Relationship }} & & \multicolumn{3}{c}{\textbf{Attribute\&Relationship }} & & \multirow{2}{*}{\textbf{Average }} \\
    \cline{2-4} \cline{6-8} \cline{10-12}
    & \textbf{SE}  &  \textbf{CE} & \textbf{Avg.} & & \textbf{SE} & \textbf{CE} & \textbf{Avg.}  & & \textbf{SE} & \textbf{CE} & \textbf{Avg.} & &\\
    \midrule
    GPT-4o & 75.58 & \textbf{88.19} & \textbf{80.64} & & 65.12 & \textbf{92.08} &80.13 & &  53.19 & \hspace{0.4cm}\textbf{83.70}& \textbf{68.28}&&\textbf{76.79}  \\
    LLaMA-3.1-Instruct-405B & 79.07 & 78.22 & 78.61 & & 77.91 & \textbf{92.08} &\textbf{85.56} & &  \textbf{57.45} & \hspace{0.4cm}71.74& 64.51&&76.25  \\
    O1-mini & \textbf{81.40}& 78.22 &79.78&&  74.42&83.17&79.14&&54.26&\hspace{0.4cm}\textbf68.48& 61.29&&73.39  \\
    Qwen-Plus & 74.42& 62.38 &67.91&&  88.37&77.23&82.35&&35.11&\hspace{0.4cm}\textbf{83.70}& 59.14&&69.82  \\
    Qwen-2.5-Instruct-72B & 75.58& 57.43 &65.78&&  89.53&78.22&83.42&&32.98&\hspace{0.4cm}78.26& 55.38&&68.21  \\
    Claude-3.5-Sonnet & 72.09& 69.31 &70.59&&  68.60&77.23&73.36&&\textbf{57.45}&\hspace{0.4cm}42.39& 50.00&&64.64  \\
    DeepSeek-V2.5 & 55.81& 60.40 &58.29&&  81.40&72.28&76.47&&42.55&\hspace{0.4cm}51.09& 50.00&&61.61  \\
    LLaMA-3.1-Instruct-8B & 44.19& 26.73 &34.76&& 58.14&18.81&36.90&&31.91&\hspace{0.4cm}21.74& 26.88&&32.86  \\
    Mistral-7B-Instruct-V0.3 & 27.90& 9.90 &18.18&& 44.19&10.89&26.20&&18.09&\hspace{0.4cm}15.22& 16.67&&20.36  \\
    \bottomrule
  \end{tabular}%
}

\label{table:base_results}
\end{table*}

\section{EXPERIMENTS}
\label{sec:experiments}
In this section, we benchmark various open-source and closed-source models using \textit{AmbiGraph-Eval}, providing a comparative analysis of their performance. Cypher is used as the target language for all experiments.
\subsection{LLM Models}
\label{sec:models}
To comprehensively analyze the different ambiguities of various competitive LLMs in \textit{AmbiGraph-Eval}, we evaluate 4 closed-source LLMs, including GPT-O1-mini \citep{openai2024o1mini}, GPT-4 \citep{openai2023gpt4}, Claude-3.5-Sonnet \citep{claude2024claude3.5}, Qwen-Plus \citep{Tongyi2024qwen}.  We also evaluate 4 open-source LLMs, including Qwen-2.5 \citep{Tongyi2024qwen}, LLaMA-3.1 \citep{meta2024llama3.1}, DeepSeek-V2.5 \citep{deepseek2024deepseekv2.5}, Mistral-7B-V0.3 \citep{mistral2023mistral-7b}. The evaluation is conducted by API calls or 4x NVIDIA A40 GPUs.

\subsection{Metrics}
\label{sec:metrics}
In the text-to-CQL task, the goal is to generate Cypher Query Language (CQL) code that not only compiles successfully but also retrieves the correct results from the database. Traditional static metrics, such as syntax match, assess syntax and structure but do not capture execution performance. Code can score well on these metrics yet fail during compilation or execution. Therefore, we focus on metrics that verify whether the generated CQL code compiles and executes as expected.

Ambiguity in database queries can affect result accuracy. To evaluate how well models handle ambiguity, we introduce \textbf{Ambiguity-Resolved Execution Accuracy (AREA)}. When ambiguity is detected, the model must generate multiple Cypher queries, each retrieving a single correct result. 

Let the target result be \( R_{\text{target}} = \{ t_1, t_2, \dots, t_k \} \), and the model-generated results be \( R(C_1), R(C_2), \dots, R(C_n) \), where \( C_1, C_2, \dots, C_n \) are the Cypher queries generated by the model. The validity of each query is determined by the function \( g \):

{\small
\[
g(R(C_i)) = 
\begin{cases} 
1 & \text{if } |R(C_i)| = 1 \text{ and } R(C_i) \subseteq R_{\text{target}} \\
0 & \text{otherwise}
\end{cases}
\]
}
A query is valid only if it returns a single result within \( R_{\text{target}} \). A set coverage function \( h \) is introduced to check if the combined results cover the entire target set:
{\small
\[
h\left(\{R(C_i)\}_{i=1}^{n}, R_{\text{target}}\right) = 
\begin{cases} 
1 & \text{if } \bigcup_{i=1}^{n} R(C_i) = R_{\text{target}} \\
0 & \text{otherwise}
\end{cases}
\]
}

The metric AREA for the task is defined as:
{\small
\[
AREA = h(\{R(C_1), R(C_2), \dots, R(C_n)\}, R_{\text{target}})
\]
}

For the entire task set, the percentage of successful tasks is calculated as:
{\small
\[
P_{\text{AREA}} = \frac{1}{N}\sum_{i=1}^{N} AREA_i
\]
}
To our knowledge, AREA is the first metric that accurately reflects the ability of LLMs to handle ambiguity in graphs during text-to-CQL through actual execution tests.

\subsection{Results \& Analysis}
\label{sec:result and analysis}

In Table \ref{table:base_results}, we evaluate the zero-shot performance on our \textit{AmbiGraph-Eval} Benchmark. The findings reveal substantial performance disparities among models in resolving graph data ambiguities. GPT-4o consistently demonstrated the best overall performance, excelling in both attribute and relationship ambiguity tasks. In contrast, while O1-mini delivered balanced results across most tasks, it was surpassed by GPT-4o in handling ambiguities requiring a deeper understanding of language and data structures. O1-mini, though adept at reasoning in tasks like coding or mathematical problem-solving, did not perform as well in this benchmark. Additionally, some models displayed task-specific strengths but struggled with distinguishing between same-entity and cross-entity scenarios.

\textbf{Attribute ambiguity}. O1-mini achieves the highest performance in same-entity (SE) attribute ambiguity tasks, with GPT-4o and LLaMA-3.1 closely following. These models show strong abilities in resolving ambiguities related to attributes within a single entity. However, in cross-entity (CE) attribute ambiguity tasks, GPT-4o significantly outperforms all models, showcasing superior reasoning capabilities in comparing attributes across different entities. LLaMA-3.1 and O1-mini also perform well on CE tasks, though not as effectively as GPT-4o. GPT-4o and O1-mini both maintain consistent performance across SE and CE tasks, while models like Qwen-2.5 show competency in SE tasks but struggle with CE tasks, suggesting difficulties in managing cross-entity ambiguity.

\textbf{Relationship ambiguity.} LLaMA-3.1 excels in handling relationship ambiguity tasks. Although GPT-4o shows the strongest overall performance, it shows relative weaknesses in addressing SE relationship ambiguity, indicating limitations in managing ambiguity within relationships involving the same entity. Conversely, in CE relationship ambiguity tasks, GPT-4o and LLaMA-3.1 achieve the highest scores, highlighting their strong abilities to handle CE ambiguity. These models particularly excel when comparing relationships across different entities, effectively tackling the complex challenges posed by such ambiguities.

\textbf{Attribute-relationship ambiguity.} LLaMA-3.1 excels in SE attribute-relationship ambiguity tasks, demonstrating stability when resolving multi-dimensional ambiguities within the same entity. This reveals that these models face challenges in addressing attribute-relationship ambiguities within single entities. In CE tasks, GPT-4o continues to perform well, leading in cross-entity ambiguity challenges, with Qwen-Plus also showing notable strength. Overall, attribute-relationship ambiguity proves to be the most challenging category, with all models generally performing worse compared to handling attribute or relationship ambiguities independently.

\section{CONCLUSION and FUTURE WORK}
In this work, we introduce \textit{AmbiGraph-Eval},, a novel benchmark for evaluating the capability of large language models (LLMs) to resolve ambiguity in graph database queries. We assess nine widely-used LLMs—both proprietary and open-source—on this benchmark. The results underscore the difficulty of handling ambiguous queries: many models fail to consistently produce correct Cypher statements. Notably, our analysis shows that strong general reasoning abilities do not necessarily translate into better performance in this task, highlighting the need for a more nuanced understanding of both ambiguity and graph query syntax.

Our evaluation identifies several major challenges that current models face on AmbiGraph-Eval: detecting ambiguous intent, generating syntactically valid Cypher Query Language (CQL), interpreting task instructions and graph structures, performing numerical aggregations, and correctly prioritizing comparisons. Among these, ambiguity detection and query syntax generation emerge as primary bottlenecks that significantly hinder overall performance.

We suggest that future research focus on jointly advancing models' capabilities in ambiguity resolution and syntax generation. Approaches such as syntax-aware prompting and explicit ambiguity highlighting may offer promising directions. Furthermore, extending the benchmark to include more complex scenarios—such as multi-hop reasoning and intricate relational patterns—could expose additional limitations of current models and guide the development of more robust solutions.

\bibliographystyle{ACM-Reference-Format}
\bibliography{sample-base}

\end{document}